\documentstyle[12pt]{article}

\input epsf
\begin{document}
\baselineskip=20.5pt

\def\beqra{\begin{eqnarray}} \def\eeqra{\end{eqnarray}}
\def\beqast{\begin{eqnarray*}}
\def\eeqast{\end{eqnarray*}}
\def\beq{\begin{equation}}      \def\eeq{\end{equation}}
\def\be{\begin{enumerate}}   \def\ee{\end{enumerate}}

\def\fnote#1#2{\begingroup\def\thefootnote{#1}\footnote{
#2}
\addtocounter
{footnote}{-1}\endgroup}

\def\itp#1#2{\hfill{NSF-ITP-{#1}-{#2}}}

\def\gam{\gamma}
\def\Gam{\Gamma}
\def\la{\lambda}
\def\eps{\epsilon}
\def\La{\Lambda}
\def\si{\sigma}
\def\Si{\Sigma}
\def\al{\alpha}
\def\Tha{\Theta}
\def\tha{\theta}
\def\vphi{\varphi}
\def\del{\delta}
\def\Del{\Delta}
\def\ab{\alpha\beta}
\def\om{\omega}
\def\Om{\Omega}
\def\mn{\mu\nu}
\def\mun{^{\mu}{}_{\nu}}
\def\kap{\kappa}
\def\rsi{\rho\sigma}
\def\beal{\beta\alpha}

\def\til{\widetilde}
\def\htil{\tilde{H}}
\def\rta{\rightarrow}
\def\eqv{\equiv}
\def\nab{\nabla}
\def\pa{\partial}
\def\sit{\tilde\sigma}
\def\ul{\underline}
\def\indt{\parindent2.5em}
\def\nd{\noindent}

\def\rsi{\rho\sigma}
\def\beal{\beta\alpha}

\def\caa{{\cal A}}
\def\cb{{\cal B}}
\def\cac{{\cal C}}
\def\cd{{\cal D}}
\def\ce{{\cal E}}
\def\cf{{\cal F}}
\def\cg{{\cal G}}
\def\cah{{\cal H}}
\def\ci{{\cal I}}
\def\cj{{\cal{J}}}
\def\ck{{\cal K}}
\def\cl{{\cal L}}
\def\cm{{\cal M}}
\def\cn{{\cal N}}
\def\cO{{\cal O}}
\def\cp{{\cal P}}
\def\car{{\cal R}}
\def\cs{{\cal S}}
\def\ct{{\cal{T}}}
\def\cu{{\cal{U}}}
\def\cv{{\cal{V}}}
\def\cw{{\cal{W}}}
\def\cx{{\cal{X}}}
\def\cy{{\cal{Y}}}
\def\cz{{\cal{Z}}}

\def\raisenot{\raise .5mm\hbox{/}}
\def\nota{\ \hbox{{$a$}\kern-.49em\hbox{/}}}
\def\notA{\hbox{{$A$}\kern-.54em\hbox{\raisenot}}}
\def\notb{\ \hbox{{$b$}\kern-.47em\hbox{/}}}
\def\notB{\ \hbox{{$B$}\kern-.60em\hbox{\raisenot}}}
\def\notc{\ \hbox{{$c$}\kern-.45em\hbox{/}}}
\def\notd{\ \hbox{{$d$}\kern-.53em\hbox{/}}}
\def\notbd{\ \hbox{{$D$}\kern-.61em\hbox{\raisenot}}} 
\def\note{\ \hbox{{$e$}\kern-.47em\hbox{/}}}
\def\notk{\ \hbox{{$k$}\kern-.51em\hbox{/}}}
\def\notp{\ \hbox{{$p$}\kern-.43em\hbox{/}}}
\def\notq{\ \hbox{{$q$}\kern-.47em\hbox{/}}}
\def\notW{\ \hbox{{$W$}\kern-.75em\hbox{\raisenot}}}
\def\notz{\ \hbox{{$Z$}\kern-.61em\hbox{\raisenot}}}
\def\notpa{\hbox{{$\partial$}\kern-.54em\hbox{\raisenot}}}

\def\fo{\hbox{{1}\kern-.25em\hbox{l}}}  
\def\rf#1{$^{#1}$}
\def\bx{\Box}
\def\tr{{\rm Tr}}
\def\rmtr{{\rm tr}}
\def\dgg{\dagger}

\def\lag{\langle}
\def\rag{\rangle}
\def\bmid{\big|}
\def\pw{P\left(w\right)}

\def\vlap{\overrightarrow{\La p}} 
\def\lrta{\longrightarrow}
\def\lrar{\raisebox{.8ex}{$\longrightarrow$}}
\def\rlarw{\longleftarrow\!\!\!\!\!\!\!\!\!\!\!\lrar}

\def\llra{\relbar\joinrel\longrightarrow}     
\def\mapright#1{\smash{\mathop{\llra}\limits_{#1}}}
\def\mapup#1{\smash{\mathop{\llra}\limits^{#1}}}
\def\asymptotic{{_{\stackrel{\displaystyle\longrightarrow}
{x\rightarrow\pm\infty}}\,\, }} 
\def\asymptext{\raisebox{.6ex}{${_{\stackrel{\displaystyle\longrightarrow}
{x\rightarrow\pm\infty}}\,\, }$}} 

\def\7#1#2{\mathop{\null#2}\limits^{#1}}   
\def\5#1#2{\mathop{\null#2}\limits_{#1}}   
\def\too#1{\stackrel{#1}{\to}}
\def\tooo#1{\stackrel{#1}{\longleftarrow}}
\def\nout{{\rm in \atop out}}

\def\one{\raisebox{.5ex}{1}}
\def\BM#1{\mbox{\boldmath{$#1$}}}

\def\ltsim{\matrix{<\cr\noalign{\vskip-7pt}\sim\cr}}
\def\gtsim{\matrix{>\cr\noalign{\vskip-7pt}\sim\cr}}
\def\haf{\frac{1}{2}}

\def\hrta{\hookrightarrow}


\def\place#1#2#3{\vbox to0pt{\kern-\parskip\kern-7pt
                             \kern-#2truein\hbox{\kern#1truein #3}
                             \vss}\nointerlineskip}

\def\illustration #1 by #2 (#3){\vbox to #2{\hrule width #1
height 0pt
depth
0pt
                                       \vfill\special{illustration #3}}}

\def\scaledillustration #1 by #2 (#3 scaled #4){{\dimen0=#1
\dimen1=#2
           \divide\dimen0 by 1000 \multiply\dimen0 by #4
            \divide\dimen1 by 1000 \multiply\dimen1 by #4
            \illustration \dimen0 by \dimen1 (#3 scaled #4)}}

\def\ON{{\cal O}(N)}
\def\UN{{\cal U}(N)}
\def\bdPh{\mbox{\boldmath{$\dot{\!\Phi}$}}}
\def\bPh{\mbox{\boldmath{$\Phi$}}}
\def\bPhs{\bPh^2}
\def\sef{S_{eff}[\sigma,\pi]}
\def\sigx{\sigma(x)}
\def\pix{\pi(x)}
\def\bph{\mbox{\boldmath{$\phi$}}}
\def\bphs{\bph^2}
\def\ex{\BM{x}}
\def\exs{\ex^2}
\def\xdot{\dot{\!\ex}}
\def\y{\BM{y}}
\def\ys{\y^2}
\def\ydot{\dot{\!\y}}
\def\pat{\pa_t}
\def\pax{\pa_x}

\renewcommand{\theequation}{\arabic{equation}}


\vspace*{.3in}
\begin{center}
 \large{\bf Taniguchi Lecture on Principal Bundles on Elliptic Fibrations}

\vspace{36pt}

Ron Donagi \noindent \footnote{email:  donagi@math.upenn.edu\\ 
Partially supported by NSF grant DMS 95-03249 and 
(while visiting ITP) by NSF grant
PHY94-07194.}  \\ \vskip 1.2cm

\end{center}

\vskip 1mm
\begin{center}
{Department of Mathematics,}\\
{University of Pennsylvania, Philadelphia, PA 19104-6395, USA}
\vspace{.6cm}

\end{center}

\begin{minipage}{5.3in}
{\abstract~~~~~}
In this talk we discuss the description of the moduli space of 
principal G-bundles on an elliptic fibration $X \rta S$ in terms of 
cameral covers and their distinguished Prym varieties. We emphasize
the close relationship between this problem and the integrability 
of Hitchin's system and its generalizations. The discussion roughly 
parallels that of [D2], but additional examples are included and some 
important steps of the argument are illustrated. Some of the applications
to heterotic/F-theory duality were described in the accompanying ICMP 
talk.

\end{minipage}

\vspace{48pt}
\vfill
\pagebreak

\setcounter{page}{1}

\section{Introduction}

\subsection{The Question}

Consider an elliptic fibration $\pi : X \rta S$ with section 
$\sigma : S \rta X$. We are interested in
describing $\cm^G_X$, the moduli space of principal G-bundles on $X$. 

Here $G$ could be any complex
reductive group. The case relevant to string theory is when $X$ is a 
Calabi-Yau
manifold of dimension $=~ 2,3,4$, and $G ~=~ E8 ~\times~ E8$, or
Spin$(32)/Z_2$, or a subgroup of either. We will divide 
the answer into three steps.

\subsection{Steps:}
\begin{itemize}
\item Describe $\cm^G_E, ~E ~=$ elliptic curve. This is the special case when
the base $S$ is a point.  We will discuss this case in section 2.
\item Put the individual moduli spaces into a family 
$\cm^G_{X/S} ~\rta ~S$ with fiber $\cm^G_{E_s}$ over the elliptic
curve $E_S$, $s \in S$. Describe the space of sections $\Gamma(S, \cm^G_{X/S})$.
\item By sending a bundle to the family of its restrictions to
fibers, we get a fibration: 
$$\cm^G_X ~\rta~ \Gamma\left(S, \cm^G_{X/S}\right)$$
sending 
$$P ~\mapsto ~ \left(s \mapsto P_s := P {\vert}_{E_s}\right).$$ 
The remaining issue then is to describe the fibers. This will be done in 
section 3.
\end{itemize}

\subsection{The Answer}
\begin{itemize}
\item $\cm^G_E$ can be identified with the quotient $\cm^T_E/W$, where $T$ is 
the maximal torus in $G$ and $W$ is its Weyl group. $\cm^T_E$, the moduli space 
of $T$-bundles of degree $0$ on $E$, is an abelian variety, isomorphic to $E^r$ 
($r$ is the rank of $G$), or more canonically to $Hom(\Lambda,E)$, where 
$\Lambda:=Hom(T,{\bf C}^*)$ is the character lattice of $G$.
$\cm^G_E$ itself is a weighted projective space. This was proved originally 
by Looijenga [L1,L2] and also by Bernstein and Shvartsman [BS]. It was reproved 
and interpreted recently by Friedman, Morgan and Witten [FMW].
\item $\Gamma\left(S, \cm^G_{X/S}\right)$ is another weighted projective space, 
since $\cm^G_{X/S} \rta S$ is a weighted projectivization of a vector bundle. 
For groups other than $E8$, this was proved by Wirthmuller [W], and closely related 
results were obtained by K. Saito [S]. The situation for $E8$ is not known.
The crucial observation for us is that, for any $G$, the base 
$\Gamma\left(S, \cm^G_{X/S}\right)$ parametrizes a family of 
$W$-Galois covers $\widetilde{S} \rta S$ which we call {\em cameral covers}:
The cover corresponding to a section $S \rta \cm^G_{X/S}$ is the pullback of 
the cover $\cm^T_{X/S} \rta \cm^G_{X/S}$.
\item The main result of [D3] is that the fiber of $\cm^G_X$ over the point 
of $\Gamma\left(S, \cm^G_{X/S}\right)$  
corresponding to a cover $\widetilde{S} \rta S$ can be identified with the 
{\em distinguished Prym} variety $Prym_{\Lambda}(\widetilde{S})$ introduced 
in [D1] . When $\widetilde{S}$ 
is non singular, this is a product of an abelian variety and a finite group. It
can be described as the kernel of a homomorphism from 
$Hom_W(\Lambda,Pic(\widetilde{S}))$ to the finite group $H^2(W,\Lambda)$.  
The identification of the fiber with 
this group is non-canonical; the fiber is really a non-trivial torser over it.

A different description of the fiber is available in case $G$ is of type $E_n$.
In this case, as proposed in [K] and in [FMW], the cameral cover can be replaced 
by a fibration 
$U \rta S$ whose fibers are del Pezzo surfaces. (For more information on del Pezzo 
surfaces, compare [De].) In [CD] it is shown that the fiber can then be 
reinterpreted as the relative Deligne cohomology group ${\cal D}(U/S)$, whose 
connected component is the relative intermediate Jacobian $J_3(U/S)$. (A similar 
result had been proved earlier by Kanev [K] for the case that the base is 
${\bf P}^1$ and the structure group is $E_n,~~n \leq 7$. A similar result over 
${\bf P}^1$ but allowing $G=E_8$ is announced in [FMW2].)
\end{itemize}

\subsection{An Analogy}
Before proceeding, we sketch an analogous problem whose solution [D2] provides the
motivation for our approach to the present question, as well as being one of
the key ingredients in its solution. In brief, we want to think of a principal
$G$-bundle on the elliptic fibration $\pi : X \rta S$ as a kind of 
$G$-Higgs bundle on $S$ ``taking its values in the fibers".

The problem is to describe the moduli space
of ``$K$-valued principal $G$-Higgs bundles" on our base $S$:

$${\rm Higgs}^G_{S,K} :=  \{(P,\phi) \mid 
\begin{array}{l}
{P:{\rm a ~principal ~G-bundle~on}~ S}\\
{\phi \in \Gamma(S, ad P \otimes K)} 
\end{array} \} $$

\noindent Here $K$ is any line bundle on $S$. In case $S$ is a curve and $K$ 
is the canonical bundle, this is the total space of Hitchin's
integrable system. In general there is a natural map which sends a Higgs bundle to 
the collection of values of
the basic $G$-invariant polynomials $f_i,~i=1, ~\ldots, ~r:={\rm rank}(g): $

$$ {\rm Higgs}^G_{S,K} \rta \Gamma(\oplus K^{d_i})$$
$$ (P, \phi) \mapsto ~\left(f_i(\phi)\right)^r_{i=1} ~ \in \Gamma\left(S,
\oplus^r_{i=1} ~K^{\otimes d_i}\right) $$
Here $d_i$ is the degree of $f_i$. Let $\bf t$ be the Lie algebra of $T$. 
The base $\Gamma\left(\oplus K^{d_i}\right)$ parametrizes sections of the 
bundle ${\bf t} \otimes K \rta s$. Now just as in the elliptic case, pulling
back via a section induces a $W$-Galois cameral cover $\widetilde{S} \rta S$.
So the base $\Gamma\left(\oplus K^{d_i}\right)$ 
parametrizes cameral covers $\widetilde{S} \rta S$. 
The main result of [D2] is that the fiber 
over a general $[\widetilde{S}]$ is the distinguished
ed Prym, i.e. the same subgroup
$Prym_{\Lambda}(\widetilde{S})$ of $Hom_W (\Lambda, Pic \widetilde{S}) $
encountered in the elliptic story.

Here is a heuristic explanation of this result. 
Giving a (diagonalizable) endomorphism 
$\phi$ of a vector space $V$ is equivalent to giving 
a decomposition of $V$ into 
eigenspaces, plus an assignment of an eigenvalue to 
each. Next, do this in families, 
i.e. start with a K-valued $GL(n)$-Higgs bundle 
$(P,\phi)$ over $S$.
If we make the unrealistic assumption that $\phi$ is 
diagonalizable with distinct 
eigenvalues (=regular semisimple) above each point 
$s \in S$, then these eigenvalues
fit together to form an $n$-sheeted {\em spectral cover} 
$\overline{S} \rta S$, while 
the (one-dimensional) eigenspaces form a line bundle over $\overline{S}$.
A more realistic assumption might be that $\phi$ may have repeated eigenvalues, 
but has a unique Jordan block per eigenvalue. ( Such $\phi$ are called 
{\em regular}.) In this case, the cover $\overline{S} \rta S$ swept out by the 
eigenvalues is ramified, but it still carries a bundle of eigenlines. Note
that $\overline{S}$ comes with an embedding in the total space of $K$.
The cameral cover $\widetilde{S}$ is the $n!$-sheeted cover which is the 
Galois closure of $\overline{S}$, i.e. a point of $\widetilde{S}$ is an 
ordered $n$-tuple of points in a fiber of $\overline{S} \rta S$. This looks 
``too big"; but it turns out to provide a natural and uniform way to extend the 
vector bundle description to other structure groups, without having to fix a 
representation. If we do pick an irreducible representation $\rho_\lambda$ of 
$G$, with highest weight $\lambda$, we recover the spectral cover 
$\overline{S}_\lambda$ of the associated $GL(n)$-Higgs bundle as the quotient 
of $\widetilde{S}$ by the Weyl subgroup fixing $\lambda$. (More precisely, 
$\overline{S}_\lambda$ has various components corresponding to the Weyl orbits 
of weights of $\rho_\lambda$, and each of these is a quotient of $\widetilde{S}$
by a Weyl subgroup.) The line bundles on 
$\overline{S}$ are replaced by their pullbacks to $\widetilde{S}$, where 
they can be characterized in terms of their behavior under the action of $W$.

The roles of the various ingredients become clearer when we introduce the 
notion of an abstract
{\em principal G-Higgs bundle} on $S$. This is a pair $(P_S, \cac)$ where $P_S$ 
is a principal G-bundle on $S$, and
$\cac \subset ad (P_S)$ is a vector subbundle whose fibers are centralizers 
of regular elements. We then think of a $K$-valued Higgs bundle as an abstract 
Higgs bundle $(P_S, \cac)$ plus a section $\phi \in \Gamma(\cac \otimes K)$. An 
abstract Higgs bundle corresponds to an (abstract) cameral cover 
$\widetilde{S} \rta S$ together with a point in its distinguished Prym. The 
additional data needed for a $K$-valued Higgs bundle then amounts to a 
collection of ``value maps" $v(\lambda) : \widetilde{S} \rta K$, one for each 
character $\lambda$ of our group $G$; the image of $v(\lambda)$ is the 
$\lambda$-spectral cover. These value maps are equivariant under the $W$ action 
on the $\lambda$ and on $\widetilde{S}$. 

The point of our analogy is the following. According to Atiyah [A], a semistable 
rank $n$ vector bundle $V_E$ on an elliptic curve $E$ decomposes as a sum of 
simple pieces which generically (in the ``regular semisimple  case") are 
distinct line bundles. 
We may therefore think of the vector bundle as consisting of a decomposition (into 
line subbundles) plus an assignment of a ``value" to each; but this value now lives 
in $Pic^0(E)$, which we canonically identify with $E$ itself.
(The corresponding statement for a generic {\em principal} bundle $P_E$ is simply 
that its  structure group can be reduced to $T$.)
A bundle $P_X$ on the elliptic fibration $\pi : X \rta S$ can therefore be interpreted 
as a cameral cover $\widetilde{S}$, plus a point in its distinguished Prym, plus a 
value map $v: \widetilde{S} \times \Lambda \rta X$. The first two ingredients 
specify an abstract Higgs bundle $(P_S, \cac)$ on $S$, the one obtained by restricting 
our principal
bundle $P_X$ as well as its Atiyah decomposition to $S$ (which is identified, 
via $\sigma$,
with the $0$-section of $X$); the value map tells us how to lift $(P_S, \cac)$ 
to a bundle on all of $X$, just as the value map
$v: \widetilde{S} \times \Lambda \rta K$
tells us how to lift $(P_S, \cac)$ to a $K$-valued Higgs bundle.

\pagebreak

\section{Bundles on an elliptic curve}

Our setup is as follows. $(E,0)$ is an elliptic curve, which we identify with
its dual 
$Pic^0~(E)$.  $G$ is a simply connected, complex semisimple group with Lie algebra
{\bf g}. (Most of what we say extends to
reductive $G$.) $B$ is a Borel subgroup, $T$ a maximal torus, $N:=N_G(T)$ the 
normalizer of $T$ in $G$, $W :=N/T$ the Weyl group, 
$\Lambda := Hom(T,C^\ast)$ the lattice of characters of $G$ (i.e. of $T$),
and $r$ = rank $G=dim ~T$.

We will work with $\cm^G_E$, the moduli of semistable G-bundles on E, and its
abelian analogue $\cm^T_E = Hom(\Lambda,E) \approx E^r$. Their relationship is:

$$ \cm^G_E = \cm^T_E / W .$$

Looijenga [L1,L2] showed that this quotient is in fact a weighted projective 
space. This result has recently been reproved and interpreted in [FMW]. 
Wirthmuller [W] showed that the identification is canonical in families, so we 
get a weighted projective bundle over the parameter space 
$\Gamma(S, \oplus ~K^{d_i})$ of cameral covers.
Wirthmuller's result holds for all simple groups except, unfortunately, $E8$. In
case $G=E8$ it is not known whether this holds. Friedman, Morgan, and Witten get
around this difficulty by restricting attention to fibrations $\pi : X \rta S$
whose singular fibers have nodes (but no cusps). Finer information about the
naturality of the trivialization and its modular properties has been obtained by
K. Saito [S].

\subsection{Regular vs. Semisimple}

A G-Bundle $P \rta E$ is:
\begin{itemize}
\item semisimple if $P ~\approx~ P_T ~\stackrel{T}{\times} ~G$, for some 
T-bundle $P_T$. 
\item regular if $h^0(E, ad ~P) ~=~r$.
\end{itemize}
Points of $\cm^G_E$ correspond to S-equivalence classes of bundles. The generic 
class has a unique representative, which is both regular and semisimple. Each 
class has a unique semisimple representative {\em and} a unique regular
representative.

\subsection{More of the Analogy}
An element $g \in G$ is:
\begin{itemize}
\item semisimple if it is in some conjugate of $T$, 
\item regular if dim $Z_G(g) = r.$
\end{itemize}
We illustrate these notions in case $G=SL(2)$:
\begin{itemize}
\item $\pm ~\pmatrix{1&0\cr 0&1}$ are the elements which are semisimple but not 
regular.
\item $\pm ~\pmatrix{1&\ast\cr 0&1}$ are regular but not semisimple $(\ast \ne 
0)$.
\item Regular semisimple bundles: $\cO(p) ~\oplus ~\cO(-p), ~p \ne -p \in 
E^u$.
\item Semisimple irregular: $\cO(p) ~\oplus ~\cO(p), ~~p=-p$. 
\item Regular non semisimple: $\cO(p) ~\otimes ~A, ~~p=-p$, where A is Atiyah's
bundle, given as a non-trivial extension:
\beq \label{Atiyah}
0~ \rta \cO ~\rta A ~\rta \cO~ \rta 0.
\eeq
\end{itemize}
\subsection{All Bundles} 
In order to work in families, we need to understand not only the regular or the
semisimple bundles. How can we describe {\em all} the G-bundles on $E$?

The semisimple bundles come from $T$-bundles. These are parametrized by 
$H^1(E,T)$,
over which there is a universal $T$-bundle, hence a universal ``Poincare"
semisimple G-bundle. (The ``T" in $H^1(E,T)$ is really the sheaf of holomorphic
sections of T. A more accurate notation would be $T(\cO_E)$.)

The other bundles come from C-bundles, where $C = Z_G (g)$ is a regular
centralizer, i.e. the centralizer of a regular element $g \in G$. 
Since $C$ is still abelian these are parametrized by $H^1(E,C)$,
and again there is a universal family.

As an example, let $g$ be one of the regular but non-semisimple elements in 
$G=SL(2)$. Then $C = Z_G(g)$ is isomorphic to ${\bf C} \times {\bf Z}_2$, so
$H^1(E,C)  \approx {\bf C} \times {\bf Z}_2 \times {\bf Z}_2$. When we go to 
the associated $G$-bundles, we get only the four distinct objects of 
(\ref{Atiyah}): the regular non-semisimple bundles $\cO(p) ~\otimes ~A, ~~p=-p$.
Each of these carries a one-dimensional family of inequivalent $C$-structures, 
parametrized by the extension class in (\ref{Atiyah}).

So in order to describe all bundles uniformly, 
we need to fit the maximal tori and the other regular centralizers into a 
single family. We know that the family of maximal tori is parametrized by 
the quotient $G/N$. It has a natural W-Galois cover, $G/T$, parametrizing 
pairs consisting of a maximal torus and a Borel subgroup containing it: 

\beq
\matrix{\{T \subset B\} & = & G/T \cr \downarrow & &\downarrow \cr \{T\}&= &G/N \cr}
\eeq
\\
To obtain the parameter space of regular centralizers, we embed $G/N$ in the
Grassmannian of r-dimensional subspaces of the lie algebra {\bf g}, and take an
appropriate open subset of the closure:

\beq
\matrix{\{C \subset B\} &= &\overline{G/T} \cr \downarrow & &\downarrow \cr 
\{C\}&= &\overline{G/N} \cr}
\eeq
\\

\noindent The key point is that over this base there exists a 
universal object $\cu_E$ parametrizing  the data:
\begin{quote}
\{bundles $P \rta E$ trivialized at $0 \in E$ + reduction to a regular
centralizer\}.
\end{quote}

\noindent{\bf Example: $G=SL(2)$}

$$\matrix{G/T &\hrta &\overline{G/T} \cr \downarrow & &\downarrow \cr G/N &\hrta
&\overline{G/N}} ~~~~~~=~~~~~~ \matrix{{\bf P}^1 \times {\bf P}^1 \setminus diagonal 
&\subset &{\bf P}^1 \times {\bf P}^1 \cr
\downarrow & &\downarrow \cr
{\bf P}^2 \setminus conic &\subset &{\bf P}^2 \cr}$$
\\
In this case, $\cu_E$ is obtained from $\overline{\cu_E} := ({\bf P}^1 \times {\bf P}^1
\times E) /{\bf Z}_2$.  First we resolve the singularities. This yields degenerate
fibers of type $I_0^\ast$. We then discard multiple components of fibers. 
This gives us $cu_E$ which maps onto ${\bf P}^2 = \overline{G/N}$. The off-diagonal 
fibers are isomorphic to $E \approx Pic^0(E)$, which parametrizes the 
$T={\bf C}^*$-bundles on $E$. The fibers over points of the diagonal are isomorphic to 
type $I_0^\ast$ curves with the central, multiplicity-$2$ component removed; this leaves 
four disjoint copies of ${\bf P}^1 \setminus \infty \approx {\bf C}$, which exactly 
matches our previous description of $H^1(E,C)$ in this case.

To do this for an arbitrary group, we start with 
$\overline{\cu}_E = (\overline{G/T} \times \cm_E^T)/W$
and resolve its singularities to obtain ${\cu}_E^+$. 
We let ${\cu}_E \subset {\cu}_E^+$ be the open subset obtained by removing 
the proper transform of the components with multiplicity $\geq 1$, i.e. the
components where some element of $W$ stabilizes the centralizer $C$ but does 
not stabilize the $T$-bundle $L$:
\beq
\matrix{\overline{\cu}_E &= &(\overline{G/T} \times~ \cm_E^T)/W \cr \cup & &\cup 
\cr
\overline{\cu}^{\prime}_E &= &\{(C,L) \mid Stab_W^C ~\subset~ Stab_W L\} \cr 
\uparrow \cr
\cu_E }
\eeq

\vspace{48pt}
\beq
\matrix{H^1(E,C) &\hrta &\cu_E \cr \downarrow & &\downarrow \cr
\{C\} &\in &\overline{G/N}}
\eeq

\pagebreak
\section{Fibrations}
\subsection{Regularized Bundles}
We now move on to the general case: $ ~ \pi : X \rightarrow S$ is an elliptic 
fibration with a section
$\sigma : S \rta X$. The restriction of our principal bundle $P \rta X$
gives a principal bundle $P_S \rta S$ on the base. The sheaf of automorphisms
along the fibers, $Aut_S(P) ~:=~ \pi_\ast Ad(P) ~\subset ~Ad (P_S)$, is 
a subsheaf of $Ad(P_S).$

The bundle $P_S$ determines the bundle $P_S/N$ of maximal tori in the fibers, as
well as the family $\overline{P_S/N}$ of regular centralizers in $P_S$, etc.

A section $c: S \longrightarrow ~\overline{P_{S/N}}$ determines an abelian
group scheme $\cac \rta S$, which is a subgroup scheme of $Ad(P_S)$:
\\
\beq
\matrix{\cac &\hookrightarrow &Ad(P_S) \cr
& \searrow ~ ~ ~ \swarrow & \cr  &S & }
\eeq
\\

By a {\em regularized G-bundle} on $X$ we mean a triple 
$(P, ~c : S \rta \overline{P_S/N}, ~P^\cac),$
where $\cac \subset Ad~P_S$ is the group scheme determined by $c$ as above,
$P^\cac$ is a $\pi^*\cac$-torser on $X$, and 
$P=P^{\cac} \times_{\cac}Ad~P_S$.
This amounts to a reduction of the structure group of $P$ 
to a group scheme $\cac$ of regular centralizers.
We note that if $P$ is everywhere regular then it has a unique regularization.

\subsection{Cameral Covers and Spectral Data}

{\ul {\bf A Cameral Cover}} is a W-Galois cover $\widetilde{S} \rta S$ modelled on 
$\overline{G/T} ~\rta~
\overline{G/N}$. (This means that the restriction $\widetilde{S}_0 \rta S_0$ of 
$\widetilde{S} \rta S$ to a sufficiently small open neighborhood $S_0$ of each
point $s \in S$ 
is the pullback of $\overline{G/T} ~\rta~ \overline{G/N}$ via a map of $S_0$ to
$\overline{G/N}$.) An equivalent notion is obtained if we model our covers instead on 
${\bf t} \rta {\bf t}/W$.

\vspace{72pt}
\noindent Our {\ul {\bf Spectral Data}} consists of:\\
{\bf (1)} a cameral cover $\widetilde{S}~\rta S$,\\
{\bf (2)}  a W-equivariant morphism $v : \widetilde{S}\rta\cm_{X/S}^T$ 
(or equivalently: $v' : \widetilde{S} \times \Lambda ~\rta~ X$), and\\
{\bf (3)}  a point of the distinguished Prym variety $Prym_{\Lambda}(\widetilde{S})$,
i.e. a homomorphism $\cl : \Lambda \rta Pic(\widetilde{S})$ 
(equivalently, a T-bundle on $\widetilde{S}$), 
which satisfies a certain twisted $W$-equivariance condition.\\

\noindent {\ul {\bf Remarks}} 

\noindent (1) Following the analogy of section 1.4, the map $v$ is called the 
{\em value map}. For each $\lambda \in \Lambda$, the image 
$v'(\widetilde{S} \times \lambda)$ is the {\em spectral cover} for 
the representation of $G$ of highest weight $\lambda$. Some typical examples: 
For $G=SL(n)$, there is an $n$-sheeted spectral cover 
$\overline{S} \rta S$ corresponding to the first
fundamental weight. A point of $\widetilde{S}$ is an ordering of the $n$ points 
in a fiber of $\overline{S}$. A point of the spectral cover corresponding to the 
$k$-th fundamental weight amounts to a choice of $k$ unordered points in a fiber.
For $G=E_n$, the root system can be identified with the collection of lines on a 
del Pezzo surface $dP_n$ obtained by blowing up $n$ points in ${\bf P}^2$. A 
point of the smallest spectral cover then corresponds to one of the lines on 
the surface, while a point of the cameral cover is specified by the choice of an
ordered $n$-tuple of disjoint lines. 

\noindent (2) Since $W$ acts on both $\Lambda$ and $Pic(\widetilde{S})$, we can 
consider the group ${Hom}_W(\Lambda, Pic(\widetilde{S}))$ of all $W$-equivariant 
homomorphisms. There is a natural map (cf. [D2])
$$ {Hom}_W(\Lambda, Pic(\widetilde{S}) \rta H^2(W,T),$$
and the homomorphisms $\cl$ which we allow form one coset of its kernel. 
Elements of the finite 
group $H^2(W,T)$ are given by classes of extensions of $W$ by $T$; our coset is 
the inverse image of the class $[N] \in  H^2(W,T)$ of the normalizer $N$. Actually, 
there is a further shift depending on the ramification divisor of 
$\widetilde{S}$ over $S$, but we will ignore this here. More details can be found 
in [D2]. For the purpose of this talk, we can consider the distinguished Prym as a 
black box: all we need to know is that it (together with the cameral cover) 
uniquely determines a principal G-Higgs bundle. We recall this notion which appeared 
in the introduction:

\noindent{\ul {\bf A Principal G-Higgs Bundle}} on $S$ is a pair $(P_S, \cac)$ 
where $P_S$ is a principal G-bundle on $S$, and
$\cac \subset ad (P_S)$ is a family of regular centralizers.

\subsection{The Main Result}

{\bf Theorem:} There is a natural equivalence between regularized G-bundles \\
\noindent $(P, ~c : S \rta \overline{P_S/N}, ~P^\cac)$ on an elliptic
fibration $\pi: X \rta S$, and  spectral data $(\widetilde{S}~\rta S, v, \cl)$.

\vspace{48pt}

\noindent \ul{\bf{Sketch of the proof}}

The regularized bundle $(P, ~c : S \rta \overline{P_S/N}, ~P^\cac)$
determines by restriction to the base a principal $G$-Higgs bundle
$(P_S, \cac)$. The latter is equivalent, by [D2] (``integrability of the 
Hitchin system"), to items {\bf (1,3)} of the spectral data. In the next subsection 
we will give a direct construction for recovering the value map (=data {\bf (2)})
from a regularized bundle. In the last subsection we illustrate, in case
$G = SL(3)$, how data {\bf (2)} allows us to lift a principal $G$-Higgs bundle
on $S$ to a regularized $G$-bundle on $X$. The idea is to use the ``universal family" 
$\cu_{X/S} \rta S$ with fibers  $\cu_E$; this includes enough data to rigidify the 
problem, resulting in the unique extension.

\vspace{48pt}

\beq \label{sketch}
\matrix{
{\rm Regularized ~ G-bundle ~ on~ X} &
\stackrel{{\rm direct~construction}}{\Longrightarrow} &
({\bf 1,2,3})\cr
\cr
{\rm forget} ~ \downarrow & & \cr
\cr
{\rm Principal} ~$G-${\rm Higgs~bundle~on~}$S$ &
\stackrel{{\rm integrability~of~Hitchin's}}{\Longleftrightarrow} & 
({{\bf 1,3}}) 
}\eeq

\vspace{48pt}

$\matrix{
\{ {\rm Embedded~cameral~covers} \} & 
= & 
\Gamma(S, \cm_{X/S}^G) & 
\Leftrightarrow & 
({\bf 1,2})
\cr
\cr
Prym_{\Lambda}(\widetilde{S}) &
\approx &
Hom_W (\Lambda, Pic ~\widetilde{S} &
\Leftrightarrow &
{\bf (3)}
}$

\vspace{48pt}

\subsection{ From a regularized $G$-Bundle to a value map.}

The fiber of the downward arrow in (\ref{sketch}) is the space $\cm^{\cac}_E$
of all $\cac$-torsers $P^{\cac}$ on $X$ with given $\cac \subset Ad~P_S$. 
We need to show that this is isomorphic to the space of data {\bf (2)}, i.e.  
of $W$-equivariant maps 
$v : \widetilde{S}\rta\cm_{X/S}^T$. This can be checked point-by-point in 
the base $S$,  but we have to pay close attention to the scheme structure 
of $\widetilde{S}$. Take $S$ to be a point, so $X=E$. Let $(G/B)^C$
be the {\em subscheme} of $G/B$ parametrizing Borel subgroups which contain
the given regular centralizer $C$. The claim is:

\beq\label{claim}
\cm_E^C ~\approx ~ Maps_W ((G/B)^C, \cm^T_E).
\eeq
First, we construct the morphism
\beq\label{morph}
\cm^C_E ~\times ~ \left(G/B\right)^C ~\rta ~\cm^T_E.
\eeq
Set-theoretically, this sends:
\beq
\left(P^C, B\right) ~\mapsto~\left(P^C ~\times^C ~B\right) ~\times^B ~T.
\eeq
To do this scheme theoretically, note that over $(G/B)^C$ there are bundles \\
$\cac ~\hrta ~\cb \rta \ct$:
\begin{itemize}
\item $\cb$ = restriction of the universal B-bundle on $G/B$
\item $\cac$ = the trivial C-bundle
\item $\ct := \cb/ [\cb, \cb]$, the trivial T-bundle
\end{itemize}

We see that a $C$-bundle $P^C$ on $E$ extends to a $\cac$-bundle $P^{\cac}$
on $E \times (G/B)^C$. This in turn induces bundles $P^{\cb}$ and $P^{\ct}$. 
The latter is a $T$-bundle on $E \times (G/B)^C$, so it is given by a 
classifying morphism $(G/B)^C \rta \cm^T_E$, which is what we wanted for a fixed 
$C$-bundle $P^C$. The existence of the Poincare bundle on $\cm^C_E$ implies that 
this globalizes to a morphism
$\cm^C_E ~\times ~ \left(G/B\right)^C ~\rta ~\cm^T_E$, as claimed.

Now $W$ acts on $(G/B)^C$ and on $\cm^T_E$, and we need to show that the map 
(\ref{morph}) is 
equivariant for these actions. This is actually easier to see in a global setup:
Letting $C$ vary over $\overline{G/N}$, the two sides of (\ref{morph}) become:
\beq
\matrix{\widetilde{\cu_E} ~:=~ \cu_E ~\times_{\overline{G/N}} ~\overline{G/T} 
~\rta
~\cm^T_E ~\times~\overline{G/N} \cr
(W {\rm ~acts~ \stackrel{\nwarrow ~~\nearrow}{here}})}
\eeq
We put together the following facts:
\begin{itemize}
\item $W$ equivariance is clear over $G/T$, since $(G/B)^C$ is a $W$-torser there,
so both sides of (\ref{claim}) equal $\cm_E^T$;
\item $\cu_E$ is smooth (of dimension r) over $\overline{G/N}$, which 
implies that:
\item $\widetilde{\cu_E}$ is smooth (of dimension r) over $\overline{G/T}$. Therefore:
\item the inverse image of $G/T$ is dense,
\end{itemize}
so we conclude that W-equivariance holds everywhere.

\subsection{Inversion}

Now that we have constructed a map from the left hand side of (\ref{claim})
to the right hand side, we need to show that it is an isomorphism. (Our 
discussion here is based on the forthcoming manuscript [DG].) To that end, 
we decompose our regular centralizer $C$:
\beq
C ~=~C_{ss} ~ \times ~C_{unip}
\eeq
into its semisimple part (a torus) and unipotent part (everything else: a vector 
group and a finite abelian group of components). The map in (\ref{claim}) 
decomposes:
$$
\matrix{\cm^C_E &= &\cm_{E}^{C_{ss}} &\times &\cm^{C_{unip}}_E \cr
\downarrow & &\downarrow & &\downarrow \cr
Maps_W \left((G/B)^C,\cm^T_E\right) &=
&Maps_W \left((G/B)^C_{red},\cm^T_E \right) &\times 
&Maps_W\left((G/B)^C_{conn},\cm^T_E\right)}.
$$
Here $(G/B)^C_{red}$ is the reduced structure underlying the finite scheme
$(G/B)^C$, while $(G/B)^C_{conn}$ is one of its connected components, a
one-pointed scheme. We can analyze the two pieces separately:

\noindent \begin{itemize}
\item The semisimple part: both sides are isomorphic to the invariant locus 
$(\cm^T_E)^{W_0}$, where $W_0$ is the subgroup of $W$ stabilizing a point of
$(G/B)^C_{red}$.
\item The unipotent part: both terms are $\cm^{C_{unip}}_E ~=~ C_{unip}$. In 
particular, this part of the answer is independent of the elliptic curve $E$. 
\end{itemize}

\noindent Combining these, we see that (\ref{claim}) is indeed an isomorphism 
as we claimed.\\

The following example illustrates in some detail the contributions of the toral, 
vector and finite parts of $C$ for the three types of regular centralizers in 
$G=SL(3)$. We describe each regular centralizer, then the moduli space $\cm^C_E$,
the coordinate ring $\cO((G/B)^C)$ of $(G/B)^C$, and finally 
$Maps_W((G/B)^C, \cm^T_E)$.
\vspace{24pt}

\noindent {\bf Example: $G=SL(3)$}\\

\noindent$\begin{array}{ccccccc}
\hline
C & &\cm^C_E & &\cO((G/B)^C) & &Maps_W((G/B)^C, \cm^T_E) \\ 
\hline \\
\pmatrix{\ast && \cr &\ast & & \cr &&\ast} & &\Lambda \otimes E \approx E^2 
& &{\bf C}[W]
& &\Lambda\otimes E \approx E^2 \\
\\
\pmatrix{a &b & \cr &a \cr & &a^{-2}} & &{\bf C} \times E & & {\bf C}[\eps]/\eps^2 
\times {\bf C}[W/Z_2]
 & &{\bf C}
\times E \\
\\
\begin{array}{c}\pmatrix{\omega &a &b \cr &\omega &a \cr & &\omega} \\ {\omega}^3 = 1\end{array}
& &{\bf C}^2 \times E[3]
& &\begin{array}{c}{\bf C}[x,y,z]/(\sigma_i = 0) \\ \sigma_i = \sigma_i(x,y,z) \\ 
i =1,2,3 \end{array} & &{C}^2 \times E[3]
\end{array}
$

\vspace{48pt}

{\bf Acknowledgements}~~~
It is a pleasure to thank Joseph Bernstein, Pierre Deligne, Dennis Gaitsgory, 
Eduard Looijenga, Tony Pantev, and Edward Witten for useful discussions on matters 
related to this talk. It is also my pleasure to thank 
Professors Saito, Shimizu and Ueno for organizing a splendid conference, and the 
Taniguchi Foundation for its financial support.

\newpage

{\bf References}

\noindent [A] M. Atiyah, Vector bundles over an elliptic curve, Proc. LMS 7 (1957), 414.

\noindent [BS] J. Bernstein and O.V. Shvartsman, {\it Chevalley's theorem for complex 
crystallographic Coxeter groups}, Func. Anal. Appl. 12 (1978), 308.

\noindent [CD] G. Curio and R. Y. Donagi, {\it Moduli in N=1 heterotic/F-theory duality}, 
hep-th/9801057.

\noindent [De] M. Demazure, {\it Surfaces de del Pezzo}, Seminaire sur les Singularities
des Surfaces, Lecture Notes in Mathematics vol 777, Springer-Verlag, 1980.

\noindent [D1] R.Y. Donagi, {\it Decomposition of Spectral covers}, 
Journees de Geometrie Algebrique d'Orsay, Asterisque 218 (1993) 145.

\noindent [D2] R.Y. Donagi, {\it  Spectral covers, in: Current topics in complex algebraic 
geometry}, MSRI pub. {\bf 28} (1992), 65-86, alg-geom 9505009.

\noindent [D3] R.Y. Donagi, {\it Principal bundles on elliptic fibrations}, 
Asian J. Math. {\bf1} (1997), 214-223, alg-geom/9702002.

\noindent [DG] R.Y. Donagi and D. Gaitsgory, in preparation.

\noindent [FMW] R. Friedman, J. Morgan and E. Witten, {\it Vector Bundles and 
F-Theory}, Commun. Math. Phys. {\bf 187} (1997) 679, hep-th/9701162.

\noindent [FMW2]R. Friedman, J. Morgan and E. Witten, {\it Principal G-Bundles over
elliptic curves}, alg-geom/9707004.

\noindent [K] V. Kanev, {\it Intermediate Jacobians and Chow groups of threefolds with a 
pencil of del Pezzo surfaces}, Annali di Matematica pura ed applicata (IV), Vol. 
CLIV (1989) 13.

\noindent [L1] E. Looijenga, {\it Root systems and elliptic curves}, Inv. Math. 38 
(1976), 17-32.

\noindent [L2] E. Looijenga, {\it Invariant theory for generalized root systems}, 
Inv. Math. 61 (1980), 1-32.

\noindent [S] K. Saito, {\it Extended affine root systems}, Publ. RIMS Kyoto, 
I: 21(1985),75, and II: 26(1990), 15.

\noindent [W] K. Wirthmuller, {\it Root systems and Jacobi forms}, Comp. Math. 82 (1992), 
293.

\end{document}